\documentstyle[12pt,epsf]{article}
\begin{document}
\baselineskip 16pt plus 2pt minus 2pt

\newcommand{\sst}[1]{{\scriptscriptstyle #1}}
\newcommand{\beq}{\begin{equation}}
\newcommand{\eeq}{\end{equation}}
\newcommand{\beqa}{\begin{eqnarray}}
\newcommand{\eeqa}{\end{eqnarray}}
\newcommand{\dida}[1]{/ \!\!\! #1}
\renewcommand{\Im}{\mbox{\sl{Im}}}
\renewcommand{\Re}{\mbox{\sl{Re}}}
\def\simge{\hspace*{0.2em}\raisebox{0.5ex}{$>$}
     \hspace{-0.8em}\raisebox{-0.3em}{$\sim$}\hspace*{0.2em}}
\def\simle{\hspace*{0.2em}\raisebox{0.5ex}{$<$}
     \hspace{-0.8em}\raisebox{-0.3em}{$\sim$}\hspace*{0.2em}}
\def\GAS{{G_{\sst{A}}^s}}
\def\GES{{G_{\sst{E}}^s}}
\def\GMS{{G_{\sst{M}}^s}}
\def\GAis{{G_{\sst{A}}^\sst{(I=0)}}}
\def\GEis{{G_{\sst{E}}^\sst{(I=0)}}}
\def\GMis{{G_{\sst{M}}^\sst{(I=0)}}}
\def\FIis{{F_i^\sst{(I=0)}}}
\def\FOis{{F_1^\sst{(I=0)}}}
\def\FTis{{F_2^\sst{(I=0)}}}
\def\FIS{{F_i^s}}
\def\FOS{{F_1^s}}
\def\FTS{{F_2^s}}
\def\GA{{G_{\sst{A}}}}
\def\GE{{G_{\sst{E}}}}
\def\GM{{G_{\sst{M}}}}
\def\FI{{F_i}}
\def\FO{{F_1}}
\def\FT{{F_2}}
\def\bra#1{{\langle#1\vert}}
\def\ket#1{{\vert#1\rangle}}
\def\pbar{{\bar{p}}}
\def\notder{{\not\! \partial}}
\def\mn{{m_{\sst{N}}}}
\def\mns{{m^2_{\sst{N}}}}
\def\mks{{m_{\sst{K}}^2}}
\def\mk{{m_{\sst{K}}}}
\def\fkem{{F_{\sst{K}}^{\sst{EM}}}}
\def\fks{{F_{\sst{K}}^s}}
\def\fka{{F_{\sst{K}}^a}}
\def\mpi{{m_\pi}}
\def\mpis{{m_\pi^2}}
\def\bpp{{b_1^{1/2,\,1/2}}}
\def\bpm{{b_1^{1/2,\,-1/2}}}
\def\bppm{{b_1^{1/2,\,\pm 1/2}}}
\def\bll{{b_1^{\lambda_1, \lambda_2}}}
\def\sgmus{{\bar{s}\gamma_\mu s}}
\def\Kbar{{\bar K}}
\def\sbar{{\bar s}}
\def\FOa{{F_1^{(a)}}}
\def\FTa{{F_2^{(a)}}}
\def\GEa{{G_\sst{E}^{(a)}}}
\def\GMa{{G_\sst{M}^{(a)}}}

\begin{titlepage}


\hfill{TRI-PP-98-41}

\vspace{1.0cm}

\begin{center}
{\large {\bf $K\bar{K}$-Continuum and Isoscalar Nucleon Form 
Factors}}

\vspace{1.2cm}

H.-W. Hammer$^{a,}$\footnote{email: hammer@triumf.ca} and
M.J. Ramsey-Musolf$^{b,}$\footnote{email: mjrm@phys.uconn.edu}

\vspace{0.8cm}

$^a$ TRIUMF, 4004 Wesbrook Mall, Vancouver, BC, Canada V6T 2A3\\
$^b$ Department of Physics, University of Connecticut, Storrs, CT
06269, USA\\[0.4cm]
\end{center}

\vspace{1cm}

\begin{abstract}
We analyse the isoscalar vector current form factors of the nucleon
using dispersion relations.
In addition to the usual vector meson poles, we account for the 
$K\bar{K}$-continuum contribution by drawing upon a recent 
analytic continuation of $KN$ scattering amplitudes.
For the Pauli form factor all strength in the $\phi$ region is already 
given by the continuum contribution, whereas for the Dirac form factor
additional strength in the $\phi$ region is required. The pertinent
implications for the leading strangeness moments are demonstrated as
well. We derive a reasonable range for the leading moments which is
free of assumptions about the asymptotic behavior of the form factors.
We also determine the $\phi NN$ coupling constants from the form
factor fits and directly from the $K\bar{K}\to N\bar{N}$ partial waves
and compare the resulting values. 
\\[0.3cm]
{\em PACS}: 14.20.Dh, 13.40.Gp, 11.55.Fv\\
{\em Keywords}: isoscalar nucleon form factors; dispersion relations;
strangeness 
\end{abstract}

\vspace{2cm}
\vfill
\end{titlepage}

\section{Introduction}
\label{sec:intro}
The electromagnetic form factors of the nucleon are 
fundamental quantities that parametrize the structure of the 
nucleon as revealed by virtual photons.
The understanding of these form factors is not only of importance  
in any theory or model of the strong interaction, but also serves
as ingredient for precise tests of the Standard Model, e.g. in the 
Lamb shift measurements performed recently \cite{Wei94}. 
In the past, the form factors have been extracted from
elastic electron-nucleon scattering experiments by means of the 
Rosenbluth separation.
With the advent of the new continious beam electron accelerators at 
Jefferson Lab, Bonn, Mainz, and NIKHEF, experiments
with polarized beams and/or targets have become possible.
These experiments allow for very precise measurements of those  
form factors which are suppressed in the Rosenbluth separation
(see, e.g., Ref. \cite{Dre98} and references therein). In fact,
a deviation from the well established dipole behavior has recently
been observed at Jefferson Lab \cite{Que98}. 

An essentially model independent tool 
to describe the form factors is given by dispersion theory
\cite{FGT58,Che58,DrZ60}. Based on analyticity and causality,
dispersion relations (DR) relate the real parts of the form factors  
to integrals involving their imaginary parts. The imaginary parts -- or
spectral functions -- contain information on the contributions to the
form factor dynamics made by various states in the hadronic spectrum. 
The quantum numbers of the current [$I^G(J^{PC}$)] restricts the set of
states which may contribute. For the isovector electromagnetic current
[$1^+(1^{--})$], the lowest mass states are 
$2\pi$, $4\pi$, $6\pi$, $\dots$, whereas for the isoscalar electromagnetic
current [$0^-(1^{--})$] they are $3\pi$, $5\pi$, $7\pi$, $2K$, $\ldots$
(cf. Refs. \cite{MHD97,HRM98}).

In principle, the electromagnetic spectral functions can be obtained from
experimental data. In this respect, the low-mass spectral content of the
isovector EM form factor has been well-understood for some time. 
Specifically, the contribution of the $2\pi$-continuum,
which is obtained from the electromagnetic form factor of the pion
and the reaction $N\bar{N}\to\pi\pi$, has been determined by H{\"o}hler
and Pietarinen \cite{Hoe75}. This contribution manifests both a strong
$\rho$-meson resonance as well as a pronounced un-correlated $\pi\pi$
continuum effect on the left wing of the resonance. The presence of the
un-correlated continuum implies that the isovector EM form factors cannot
be adequately represented by a simple vector meson dominance (VMD) picture
\cite{Sak69}.  Consequently, the $2\pi$-continuum has been built into the
spectral functions for the isovector nucleon form factors
\cite{Hoe76,MMD96} explicitly. The remaining strength in the isovector
channel is then parametrized by three narrow excitations of the $\rho$
meson.

The situation for the isoscalar form factors is less clear. As in the
isovector case, one expects the low-mass states to generate both
resonant and non-resonant continuum contributions. Historically, the
first analyses of these form factors assumed that the isoscalar spectral
functions can be parametrized solely by sharp vector meson resonances.
Although such an approach does not produce the correct singularity
structure for the form factors \cite{FrF60}, an adequate fit to EM
data can nevertheless be obtained \cite{Hoe76,MMD96}. 
Such fits require at least
two closely-lying resonances ({\em e.g.}, $\omega$ and $\phi$) with
opposite sign residues in order to generate the observed dipole behavior
of the form factors. A third resonance (denoted by $S'$ in Ref.
\cite{Hoe76}) is included in order to obtain an acceptable
$\chi^2$. The $S'$ effectively summarizes higher-mass spectral strength. 
One surprising implication of the VMD analysis is a large value for the 
$\phi$-nucleon coupling $g_{\phi NN}/g_{\omega NN}\approx -1/2$. This 
evidence for significant OZI-violation \cite{OZIrule} suggests large 
moments for the strange vector form factors as well \cite{Jaf89}. 

Subsequent studies have examined the validity of the VMD {\em ansatz} for
the isoscalar form factors. The authors of Ref. \cite{BKM96} computed the
un-correlated $3\pi$ continuum to leading order in chiral counting. They
find no evidence for continuum enhancement as occurs for the $2\pi$
contribution to the isovector form factors.\footnote{The multi-pion 
contributions could, however,  be  enhanced by resonance effects 
\cite{HaM98}.} The authors of Ref. \cite{MMS97} argued that the VMD 
analyses neglect important contributions from
a correlated $\rho\pi$ resonance which sits on top of the $3\pi$ continuum.
It is argued in Ref. \cite{MMS97} that the effect of the  $\rho\pi$ 
exchange can be parametrized by a single pole at $t=(1.12 \mbox{ GeV})^2$
with a residue fixed from the Bonn potential.  
A fit to isoscalar form factor data, with this effective $\rho\pi$ 
singularity included, leads to
a significantly  smaller $\phi$-nucleon coupling than obtained in Refs.
\cite{Hoe76,MMD96}. To the extent that the $\rho\pi$ resonance does not 
couple to $\bar{s}\gamma_\mu s$, one infers considerably smaller 
values for the strangeness
moments than obtained in Ref. \cite{Jaf89}. It was noted in Ref. \cite{HaM98},
however, that $\bra{\rho\pi}\bar{s}\gamma_\mu s\ket{0}$ does not vanish, 
since the $\phi$ decays to $\rho\pi$ 12\% of the time. Thus, the 
inclusion of the $3\pi\leftrightarrow \rho\pi$ resonance need not imply 
small strangeness moments.

In the wake of these analyses, several questions pertaining to the 
isoscalar EM and strangeness vector current spectral content remain:
\begin{itemize}
\item[(i)] Does any evidence exist among EM or strong interaction data for
large OZI violation in the nucleon?  
\item[(ii)] Does the VMD picture give an accurate representation
of the isoscalar EM spectral functions?
\item[(iii)] To what extent does our knowledge of the isoscalar EM spectral
functions constrain predictions for the strange quark vector current form
factors and their leading moments? 
\end{itemize}
In this paper, we address these issues by concentrating on the role of
the $K\bar{K}$-continuum. In analogy to analyses of the the 
$2\pi$-continuum for the isovector form factors,
we introduce the $K\bar{K}$-continuum into the analysis 
of the isoscalar ones and study the nature of 
the $\phi$ strength in detail. No additional parameters are introduced 
because this contribution is obtained from an analytic continuation of 
experimental $KN$ scattering amplitudes and $e^+e^-\to K\bar{K}$ 
data \cite{HRM98}.
In particular: 
\begin{itemize}
\item[(a)] We compare the $\phi NN$ couplings derived from the 
$K\bar{K}\to N\bar{N}$ partial waves with those obtained from
the original VMD analyses of the electromagnetic nucleon form factors
and comment on the validity of the OZI rule. We find that strong interaction
data imply large values for the $g_{\phi NN}$, in disagreement with the
conclusions of Ref. \cite{MMS97}. 
\item[(b)] We re-fit the isoscalar EM form factors under various scenarios
used in Refs. \cite{Hoe76,MMD96,MMS97} but also including the continuum
$K\bar{K}$ contribution explicitly. Our fits determine the phase of the
latter, which cannot be obtained from strong interaction and $e^+e^-$ data
alone, as well as the stable resonance contributions to the form factors. We
find that the $K\bar{K}$ contribution, which contains a $\phi$-resonance,
accounts for nearly all the $\phi$-strength in the isoscalar Pauli form factor,
but that additional $\phi$-strength is required in the isoscalar Dirac form
factor.
\item[(c)] Based on the analysis of (b), we argue that the the VMD approach 
represents an effective parameterization, but leads to erroneous values for the
$\phi$ nucleon couplings.  
\item[(c)] We demonstrate the pertinent implications for the nucleon's
strange vector form factors.
\end{itemize}

Our discussion of these points is organized as follows. In the next section, we
briefly  review the  necessary formalism and dispersion relations. In Section
\ref{sec:par}, we derive $\phi NN$ coupling constants from the $K\bar{K}\to 
N\bar{N}$ partial waves of Ref. \cite{HRM98}. The spectral content
of the isoscalar form factors is analysed in Section \ref{sec:spec}.
Finally, we demonstrate the consequences of our analysis for pole models 
of nucleon strangeness in Section \ref{sec:stra} and conclude in Section 
\ref{sec:conc}.

\section{Dispersion Relations}
\label{sec:form}
The vector current form factors of the nucleon, $F_1(t)$ and $F_2(t)$, 
are defined by:
\beq
\bra{N(p')} j_\mu \ket{N(p)}=\bar{u}(p')\left[
\FO(t)\gamma_\mu+{i\FT(t)\over 2\mn}\sigma_{\mu\nu}(p'-p)^\nu
\right]u(p)\, .
\eeq
where $u(p)$ is the spinor associated with the nucleon state $\ket{N(p)}$
and $t=q^2=(p'-p)^2$ is the four-momentum transfer.
We consider two cases for $j_\mu$: (i) the
strange vector current $\bar{s} \gamma_\mu s$ and (ii) the isoscalar
electromagnetic current $j_\mu^\sst{(I=0)}$.
Since the nucleon carries no net strangeness, $\FOS$ must vanish at zero
momentum transfer, whereas $\FOis$ is normalized
to the isoscalar electromagnetic charge of the nucleon, $\FOis(0)=1/2$.
Both currents couple to the same intermediate states because they
have the same quantum numbers \cite{MHD97, HRM98}.
We use a subtracted dispersion relation (DR) for $\FO$ and an unsubtracted 
one for $\FT$,
\beqa
\label{fos}
\FO(t)&=&\FO(0)+\frac{t}{\pi}\int_{9 m_\pi^2}^{\infty}
\frac{\Im\,\FO(t')}{t'(t'-t)}dt' \, ,\\
\label{fts}
\FT(t)&=&\frac{1}{\pi}\int_{9 m_\pi^2}^{\infty}
\frac{\Im\,\FT(t')}{t'-t}dt' \,.
\eeqa
The lower limit of integration is given by the threshold of the lightest
intermediate state contributing to the form factors, the $3\pi$ state.
In the VMD analyses of the isoscalar electromagnetic form factors
(see, e.g., Refs. \cite{Hoe76,MMD96}), their imaginary parts
are parametrized by narrow vector meson resonances as,\footnote{Note that 
in Refs. \cite{Hoe76,MMD96} an unsubtracted dispersion relation has been
used for $\FO$ as well.}
\begin{equation}
\Im\,  \FI(t) = \pi\sum_{j=\omega,\phi,S'} a^i_j\delta(t-m_j^2)\,.
\end{equation}
For a successful description of the data the $\omega$, the $\phi$, 
and a fictitious third $S'$ resonance at $m_{S'} = 1.6$ GeV are needed. 
Although the $S'$ can be
identified with the $\omega(1600)$ or the $\phi(1680)$ \cite{PDG98},
it effectively accounts for the strength in the high mass region.
With the DR's, Eqs. (\ref{fos}, \ref{fts}),
this leads to the usual pole parametrizations,
\beqa
\label{fosp}
\FO(t)&=& \FO(0)+\sum_{j=\omega,\phi,S'} \frac{t}{m_j^2}
\frac{a^1_j}{m_j^2-t}\,,
\\
\label{ftsp}
\FT(t)&=& \sum_{j=\omega,\phi,S'} \frac{a^2_j}{m_j^2-t}\,.
\eeqa

The residues $a^i_j$ are then fitted to the form factor data.
In contrast to the isovector nucleon form factors, no continuum
contributions are necessary to obtain successful fits. 
Based on these fits, the phenomenology of 
$\phi-\omega$-mixing, and the known flavor content of the $\omega$
and $\phi$, predictions for the strange vector form factors have 
been made \cite{Jaf89,HMD96,For96}. However, the coupling of the $S'$
to the strange vector current is uncertain. Because the flavor content of the 
$S'$ is unknown, its coupling to $\bar{s}\gamma_\mu s$ has been inferred in
Refs. \cite{Jaf89,HMD96} from {\em ad hoc}
assumptions  about the asymptotic behavior of the strangeness form factors. 
Unfortunately, the leading moments of the strange
form factors are very sensitive to the assumed asymptotic behavior 
\cite{For96,Mubos}. We present a possible solution
to this problem in Section \ref{sec:stra}. First, we focus on the 
$K\bar{K}$-continuum contribution for isoscalar nucleon form factors.

\section{$K\bar{K}$-Continuum}
\label{sec:par}
\begin{figure}[htb]
\epsfxsize=16cm
\begin{center}
\ \epsffile{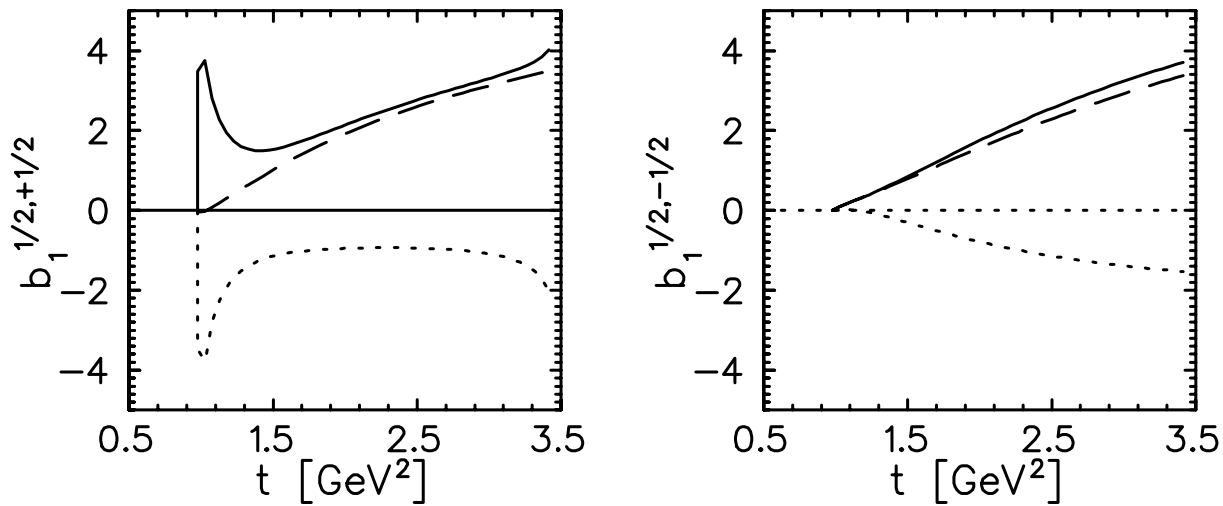}
\end{center}
\caption{\label{fig1} $\bppm$ in the unphysical region, $4\mks \leq t
\leq 4\mns$, obtained from an analytic continuation of $KN$
scattering amplitudes \cite{HRM98}.}
\end{figure}

The $K\bar{K}$ contribution to the
imaginary part of the isocalar form factors is given by \cite{MHD97,HRM98}
\begin{eqnarray}
\label{imf1}
\Im\,  \FOa(t)&=&\Re\,\left\{\left({\mn q_t\over 4 p_t^2}\right)\left[
{E\over\sqrt{2}\mn}\bpm(t)-\bpp(t)\right]\fka(t)^{\ast}\right\}\,,\\
&& \nonumber \\
\label{imf2}
\Im\,  \FTa(t)&=&\Re\,\left\{\left({\mn q_t\over 4 p_t^2}\right)\left[
\bpp(t)-{\mn\over\sqrt{2}E}\bpm(t)\right]\fka(t)^{\ast}\right\}\,,
\end{eqnarray}
with
\beq
\label{defqp}
p_t=\sqrt{t/4-\mns}\,,\quad
q_t=\sqrt{t/4-m_K^2}\,,\quad \mbox{and} \quad
E=\sqrt{t}/2\,.
\eeq
The superscript $a$ denotes $s$ or $(I=0)$ for strange and isoscalar
electromagnetic form factors, respectively. 
$\fka(t)$ represents the kaon form factor in the respective channel, 
\beq
\bra{0}j_\mu^{(a)}\ket{K(k) \bar{K}(\bar{k})}=(k-\bar{k})_\mu \fka(t)\,,
\eeq
whereas the $\bppm$ are the $J=1$ partial waves for 
$K\bar{K}\to N\bar{N}$ \cite{MHD97, HRM98}. Once these imaginary
parts are determined, the contribution of the $K\bar{K}$-continuum 
to the form factors is obtained from the DR's, Eqs. (\ref{fos}, \ref{fts}).
For $t \geq 4\mns$ the partial waves are bounded by unitarity,
\begin{equation}
\label{ubs}
|\bppm(t)|\leq 1 \,.
\end{equation}
Eq. (\ref{ubs}), however, does not hold in the unphysical region,
$4\mks \leq t \leq 4\mns$. Recently, the $\bppm$ in the unphysical region
have been determined from an analytic continuation of $KN$-scattering 
amplitudes \cite{HRM98}. The resulting partial waves are shown in 
Fig. \ref{fig1}.
The striking feature is a 
clear resonance structure at threshold in $b_1^{1/2,\,1/2}$, which
presumably is the $\phi$ resonance. However, this resonance is not seen
in $b_1^{1/2,\,-1/2}$ although it is not forbidden by
the quantum numbers of the $\phi$.
In a simple resonance model this behavior is recovered
when the vector ($g^1_{\phi NN}$) and tensor ($g^2_{\phi NN}$)
couplings of the $\phi$ meson to the nucleon are equal and have 
opposite signs. This  can be seen from the following parametrizations
\cite{BrM79,HRM98},
\beqa
\label{bphi}
\bpp (t) &=& \frac{2 q_t^2}{\sqrt{t}}\frac{2 R_+^{\phi}}{m_\phi^2-t-i
m_\phi \Gamma_\phi h_\phi(t)} \,,\\
\bpm (t) &=& q_t^2 \frac{2 R_-^{\phi}}{m_\phi^2-t-i 
m_\phi \Gamma_\phi h_\phi(t)} \,,\nonumber
\eeqa
with
\beqa
\label{phired}
R^\phi_+ &=& -\frac{2\mn}{3}\frac{g_{\phi K\bar{K}}}{4\pi}
\left(g^1_{\phi NN}+\frac{m_\phi^2}{4\mns}g^2_{\phi NN}
\right)\,,\\
R^\phi_- &=& -\frac{2\sqrt{2}}{3}\frac{g_{\phi K\bar{K}}}{4\pi}
\left(g^1_{\phi NN}+g^2_{\phi NN}\right)\,,\nonumber
\eeqa
where $\Gamma_\phi$ is the total width of the $\phi$
and $h_\phi(t)=t/m_\phi^2$ \cite{FeS81}.
The $\phi K\bar{K}$ coupling is obtained from the partial width of the 
$\phi \to K\bar{K}$ decay \cite{BrM79},
\beq
|g_{\phi K\bar{K}}| = 4.10 \pm 0.28\,.
\eeq

We have fitted the expressions from Eqs. (\ref{bphi}, \ref{phired}) 
to the absolute values of the 
amplitudes from the analytic continuation (see Fig. \ref{fig1})
in order to determine the vector and tensor $\phi NN$ couplings
$g^1_{\phi NN}$ and $g^2_{\phi NN}$, respectively. Since the $\bppm$ from
the analytic continuation contain substantial nonresonant contributions,
we only fit the region from the $K\bar{K}$ threshold to about
1.5 GeV where the $\phi$ resonance is dominating. Furthermore, we force
$g^{2}_{\phi NN}=-g^{1}_{\phi NN}$ to improve the stability of the 
fit. The resulting coupling constants are given in Table \ref{tab1}.
\begin{table}
\begin{center}~
\begin{tabular}{|c||c|c|}\hline
Scenario & $g^{1}_{\phi NN}$ & $g^{2}_{\phi NN}$
\\\hline\hline
(a) & $-7.4\pm 1.46$ & $7.4\pm 1.46$  \\\hline
(b) & $-9.6\pm 2.44 $ & $9.6\pm 2.44$ \\\hline
Ref. \protect\cite{MMD96} & $-9.16\pm 0.23$ & $2.01\pm 0.33$\\\hline
\end{tabular}
\end{center}
\caption{\label{tab1} $\phi NN$ coupling constants 
$g^{2}_{\phi NN}=-g^{1}_{\phi NN}$ as obtained from 
a fit of Eqs. (\ref{bphi}, \ref{phired}) to the partial waves from Fig.
\ref{fig1}. First two rows give our results for scenarios 
(a): $\Gamma=4.43$ MeV \protect\cite{PDG98} and (b)
$\Gamma=16\pm 10$ MeV from fit. 
Last row gives the result of Ref. \protect\cite{MMD96}.}
\end{table}
We used two scenarios for our fits: (a) the width of the $\phi$ is
taken from the particle data group \cite{PDG98} and (b) the width
is fitted together with the coupling constant. The results for both
scenarios agree within the error bars. Furthermore, $g^{1}_{\phi NN}$ 
agrees with the value obtained from the dispersion analysis of the 
electromagnetic nucleon form factors \cite{MMD96}, while
the values for $g^{2}_{\phi NN}$ are in variation. The discrepancy
in $g^{2}_{\phi NN}$ is due to our constraint $g^{2}_{\phi NN}=
-g^{1}_{\phi NN}$. When we omit this constraint, 
$g^{2}_{\phi NN}\approx 4.0$ is similar to the value of Ref. 
\cite{MMD96}, however, the quality of the fit at threshold is not
satisfactory. The size of the $\phi NN$ coupling constants in both 
scenarios is of the same order as in pole analyses and therefore
implies a large OZI-violation as well.

We believe that extracting the coupling constants from the partial 
waves rather than electromagnetic pole analyses is the more sensible procedure
for at least three reasons. First, in the VMD form factor analyses, one
effectively  summarizes the sum over a variety of intermediate state 
contributions by a few sharp resonances. The residues in this case may 
include the effects of
both true resonances as well as non-resonant continuum contributions. 

Second, the EM form factor data exist only in the space-like domain and, 
thus, do not manifest any resonance structure explicitly. 
The $KN$ partial waves, on the other hand, have been analytically 
continued into the time-like region where an isolated resonance structure 
is apparent (see Fig. \ref{fig1}). 

Finally, the
resonating contribution from a given intermediate state may not be adequately
represented by a simple VMD {\em ansatz}. To see why, consider the $K\bar{K}$
contributions to the isoscalar EM form factors, $F_i^{K\bar{K}}$. 
We have also fitted these contributions (cf. Eqs. (\ref{imf1}, 
\ref{imf2})) with an effective $\phi$ pole, 
\beqa
F_1^{K\bar{K}}(t) &=& \frac{t}{m_\phi^2}
\frac{\tilde{g}^1_{\phi NN}}{m_\phi^2-t}\frac{m_\phi^2}{f_\phi}\,,\\
F_2^{K\bar{K}}(t) &=& \frac{\tilde{g}^2_{\phi NN}
}{m_\phi^2-t}\frac{m_\phi^2}{f_\phi}\,,\nonumber
\eeqa
where $f_\phi=13$ is obtained from the width of the leptonic decay
$\phi\to e^+ e^-$. We find the following effective coupling constants:
\beq
\tilde{g}^1_{\phi NN}=1.32\pm 0.01 \quad \mbox{and} \quad
\tilde{g}^2_{\phi NN}=2.86 \pm 0.01\,.
\eeq
The effective couplings $\tilde{g}^i_{\phi NN}$ are quite
different from the $g^i_{\phi NN}$ for scenarios (a) and (b)
in Table \ref{tab1}. Consequently, we conclude that the $K\bar{K}$-continuum 
is not well represented by a vector meson dominance approximation.
Note, however, that $\tilde{g}^2_{\phi NN}$ is comparable to
$g^2_{\phi NN}$ as obtained from the pole analysis \cite{MMD96}.
The reason for this \lq\lq agreement"  will be discussed in the next section.

In deriving values for the $g^i_{\phi NN}$ from $KN$ data, we note that there
exist un-quantified theoretical uncertainties beyond those quoted in Table 1. 
A more careful treatment of the $\bll$ would have included form factors along
with simple coupling constants in Eqs. (\ref{bphi}, \ref{phired}). Ideally, 
one would want to follow the lines of Ref. 
\cite{Hoe75} where this has been done for the $2\pi$ contribution to
the nucleon's isovector form factors and the $\rho NN$ vertex. 
However, the latter treatment relies strongly on the fact that the phases of 
$\pi\pi\to N\bar{N}$ scattering and the pion 
electromagnetic form factor essentially follow
each other because the inelasticities are small.  Unfortunately, the
situation is more complicated in the kaon case, and a straightforward 
extension of the method of Ref. \cite{Hoe75} is not possible.

Hence, there is an error bar associated with our $\phi NN$
couplings due to the continuum under the peak which we 
are unable to quantify at this time.
Nevertheless, an important difference between isoscalar and the isovector 
case is that the width of the resonance peak is much smaller in the former. 
Consequently, non-resonant continuum effects  are not likely to substantially 
affect the residue for such a pronounced resonance. This expectation is 
corroborated by the fact that the coupling constants do not change 
appreciably with the width (scenario (a) vs. (b)). 

Finally, we observe that 
since the amplitudes themselves -- rather than a 
parametrization of them -- is used in Eqs. (\ref{fos}, \ref{fts}, \ref{imf1},
\ref{imf2}), the interpretation of the $\bppm$ is inconsequential 
for the following form factor analysis.

\section{Form Factor Fits}
\label{sec:spec}
We now include the $K\bar{K}$-continuum into the
dispersion analysis of the isoscalar electromagnetic nucleon form 
factors by adding it to Eqs. (\ref{fosp}, \ref{ftsp}). 
We then refit the residues of the $\omega$, $\phi$, and $S'$ poles.
The masses $m_\omega^2=0.6115 \mbox{ GeV}^2$, $m_\phi^2=1.0384 \mbox{ GeV}^2$,
and $m_{S'}^2=2.56 \mbox{ GeV}^2$ are fixed \cite{MMD96}.
Since Ref. \cite{HRM98} does not give the phase of the 
$K\bar{K}$-continuum relative to the pole contributions, we determine 
this phase from the fits as well. 
For simplicity, we do not fit to the experimental data but rather to 
the results of Ref. \cite{MMD96}.\footnote{Note that in Ref. \cite{MMD96} 
all form factors have been fitted simultaneously. Since we are
mainly interested in qualitative features, we deem this procedure here 
unnecessary.} The fits indicate a relative
phase of $0\,(\pi)$ between the $K\bar{K}$-continuum and the $\omega$
pole contribution for $\FO\,(\FT)$. 
In fact, it is impossible to obtain a satisfactory 
fit with a different relative phase between the two contributions.
After having fixed the relative phase,
we aim to determine how much of the original $\phi$ strength 
can be accounted for by the $K\bar{K}$-continuum.

We have performed fits for $\FOis$ and $\FTis$ in a number of different 
scenarios. We present the following four in detail: 
\begin{itemize}
\item[(i)]  $\omega$, $\phi$, and $S'$ poles and no $K\bar{K}$-continuum
(cf. Ref. \cite{MMD96}).
\item[(ii)] $\omega$, $\phi$, and $S'$ poles and $K\bar{K}$-continuum.
\item[(iii)] $\omega$ and $S'$ poles and $K\bar{K}$-continuum.
\item[(iv)] $\omega$, $\omega'$, and $S'$ poles and $K\bar{K}$-continuum.
\end{itemize}
The $\omega'$ pole in scenario (iv) is not physical.
It was introduced in  Ref. \cite{MMS97} to parametrize the contribution of
the $\rho\pi$-continuum obtained from the Bonn potential.
We take the same 
mass, $m_{\omega'}^2=1.2544\mbox{ GeV}^2$, as in Ref. \cite{MMS97}.
However, we let the residue free because our $K\bar{K}$-continuum is 
different from the one used in Ref. \cite{MMS97}. The results of the 
fits are shown in Table \ref{tab2}.
\begin{table}
\begin{center}~
\begin{tabular}{|c||c|c|c|c||c|c|c|c|}\hline
Scenario & $g^{1}_{\omega NN}$ & $g^{1}_{\phi NN}$ & $a^1_{S'}$ &
$a^1_{\omega'}$ & $g^{2}_{\omega NN}$ & $g^{2}_{\phi NN}$ & $a^2_{S'}$ &
$a^2_{\omega'}$ \\\hline\hline
(i) & $21.1$ & $-9.7$ & $0.0035$ & - & $-3.36$ & $1.98$ & $-0.038$ & -   
\\\hline
(ii) & $21.2$ & $-10.7$ & $-0.072$ & - & $-3.31$ & $-0.46$ & $-0.11$ & -  
\\\hline
(iii) & $13.1$ & - & $-1.08$ & - & $-3.72$ & - & $-0.142$ & -  
\\\hline
(iv) & $19.1$ & - & $0.21$ & $-1.01$ & $-3.42$ & - & $-0.104$ & $-0.04$ 
\\\hline
\end{tabular}
\end{center}
\caption{\label{tab2} Fitted residues $a_V^i$ for scenarios (i)-(iv). 
The residues $a_V^i$ are given in units of GeV$^2$.
For the $\omega$ and $\phi$ poles the couplings 
$g^i_{VNN}=\frac{f_V}{m_V^2}a_V^i$ with $f_\phi =13$ and $f_\omega =17$
are shown instead of the residue. 
}
\end{table}
Scenario (i) corresponds to the original analysis of Ref. \cite{MMD96}.
In the other scenarios the strong $\phi$ coupling demanded by the 
data is partially or fully
accounted for by the $K\bar{K}$-continuum or the $\rho\pi$-continuum
(modelled by the effective $\omega'$ pole). Most of the fits give a similarly
accurate description of the form factors from Ref. \cite{MMD96}. The fit
for scenario (iii) is somewhat poorer because there is one fitparameter
less.

We now make several observations based on the fits. In particular,  
the $\omega$ contribution is very stable in all scenarios. However, this
is not the case for the residual $\phi$ and the $S'$. Since the $S'$ is not 
a physical vector meson but effectively summarizes higher lying strength
this is neither surprising nor alarming.

To understand the role of the $\phi$, consider in detail $\FTis$: when the 
$K\bar{K}$-continuum is introduced in scenario (ii), the $\phi NN$ coupling 
is reduced considerably. In fact, in scenarios (iii) and (iv) the form 
factor can be described without a $\phi$ pole at all. Consequently, all
the $\phi$ strength can be accounted for by either the $K\bar{K}$-continuum 
alone or in tandem with a small, effective $\rho\pi$-continuum contribution. 
This is the reason why the  effective coupling coupling $\tilde{g}^2_{\phi NN}$
from the previous  section is comparable to $g^2_{\phi NN}$ from Ref.
\cite{MMD96} (even though the effective coupling  $\tilde{g}^2_{\phi NN}$ is
not the same as the tensor coupling extracted from the $KN$ partial waves). 

The situation is different, however, for $\FOis$: when the 
$K\bar{K}$-continuum is introduced in  scenario (ii), the 
$\phi NN$ coupling is almost unchanged. This suggests additional 
contributions in the $\phi$ region that are mocked by the $S'$
in scenario (i). Further evidence is provided by scenario (iii): 
removing the $\phi$ pole significantly changes the otherwise very stable 
$\omega NN$ coupling by a factor of 2 and leads to a unnaturally large
coupling for the $S'$. If the $\phi$, however, is replaced by the
$\omega'$ as in scenario (iv), reasonable couplings for the $\omega$
and $S'$ are obtained. In contrast to $\FTis$ there appears to be a 
considerable contribution from the $\rho\pi$-continuum to $\FOis$, and 
the $\phi$ can not be accounted for by the $K\bar{K}$-continuum alone. 

We conclude that the role of the $\phi$ in $\FTis$ is well understood, and 
that it cannot be represented by a simple VMD structure with the physical 
$\phi$ nucleon couplings. The role of the $\phi$ in $\FOis$ remains ambiguous. 
This ambiguity stems from the fact that
(a) the $K\bar{K}$ contribution does not saturate the spectral function 
strength in the $\phi$-region and (b) equally acceptable fits are obtained 
whether one saturates this strength either with a $\phi$ pole explicitly or 
with an effective $\rho\pi$ ($\omega'$) pole. Moreover, the flavor 
structure of the latter is also open to debate, since the $\rho\pi$ 
isoscalar EM and strangeness form factors
contain $\phi$-strength \cite{HaM98,Goi96}. We suspect, nevertheless, that 
nearly all of the strength in $t\approx 1$ GeV$^2$ region is due to the $
\phi$, since the values of $g^1_{\phi NN}$ obtained in scenarios (i) and (ii)
agree with the values obtained from the $KN$ partial waves. 

\section{Strange Moments}
\label{sec:stra}
Finally, we use the fits from above to obtain information on 
the leading strange moments,
\beq
\label{moms}
\kappa^s = \FTS(0)\,,\vphantom{\frac{1}{2}}\quad \mbox{and} \quad
\langle r^2\rangle^s_D = 6{d\FOS(t)\over d t}\bigg\vert_{t=0}\,.
\eeq
In pole models \cite{Jaf89,HMD96,For96},
the couplings of the $\phi$ and $\omega$ to the strange vector
current are inferred from their known flavor content and coupling
to the isoscalar electromagnetic current. The ratios of the 
corresponding pole residues are \cite{Jaf89, HaM98}
\begin{eqnarray}
\label{resratom}
(a_\omega^{i})^s/a^i_\omega &=& -\sqrt{6}\left[{\sin\epsilon\over\sin(
\epsilon+\theta_0)}\right]\approx -0.2\,, \\
(a_\phi^{i})^s/a^i_\phi &=& -\sqrt{6}\left[{\cos\epsilon\over\cos(\epsilon+
\theta_0)}\right]\approx -3\,, 
\label{resratphi}
\end{eqnarray}
where the superscript $s$ denotes the residue for the strangeness form 
factor, $\theta_0=0.6154$ is the \lq\lq magic" octet-singlet mixing angle 
giving rise to pure $u\bar{u}+d\bar{d}$ and $s\bar{s}$ states and 
$\epsilon=0.055$ deviations from ideal mixing.
Since the flavor content of the $S'$ is not known, its coupling to the 
strange current was fixed by an asymptotic condition. 
Here, we follow a slightly different approach. 

We consider scenarios (i) and (ii)
from Section \ref{sec:spec} corresponding to $\omega$, $\phi$, and
$S'$ poles without and with $K\bar{K}$-continuum, respectively.
The residues of the $\omega$ and $\phi$ poles have been obtained
from fits to the isoscalar form factors in the previous section. 
As in Refs.  \cite{Jaf89,HMD96,For96}, we draw upon simple
flavor rotation arguments leading to Eqs. (\ref{resratom}, 
\ref{resratphi}) to determine the corresponding residues for the
strange vector form factors. The $K\bar{K}$ contribution is also
known \cite{HRM98} and its phase has been determined from fits to the 
electromagnetic form factors. For illustrative purposes, we also consider 
scenario (iv), under the assumption that the intermediate
$3\pi\leftrightarrow\rho\pi$ state (parameterized as the $\omega'$ in 
Ref. \cite{MMS97}) does not couple to
$\bar{s}\gamma_\mu s$. Although the latter {\em ansatz} is not well-justified
phenomenologically \cite{HaM98,Goi96}, we include it to demonstrate the 
sensitivity of our predictions to rather extreme assumptions.

The combined contributions from the $\omega$-resonance, $K\bar{K}$
continuum, and residual $\phi$-strength to the strangeness moments are 
listed in Table 3 as the \lq\lq low-mass" values for the moments. These 
contributions are strongly constrained by the phenomenology of EM form 
factor data, $KN$ scattering phase shifts, $e^+e^-\to K\bar{K}$ cross 
sections, and vector meson flavor content.
It is difficult to maintain consistency with these phenomenological 
inputs and evade
the low-mass contribution to the strangeness moments given in Table 3.   
One might have expected the use of scenario (iv) -- where the $\phi$ 
strength is
replaced by the effective $\rho\pi$ contribution -- to yield smaller low-mass
values. In fact, the low-mass value for
$\kappa^s$ under scenario (iv) is similar to the other values in Table 3, 
since the effective $\rho\pi$ contribution to 
$\FTis$ is negligible and since the $K\bar{K}$ contribution
saturates the $\phi$-strength. In the case of $\langle r^2\rangle^s_D$, the
impact is potentially more significant. If one assumes the resonating
$\rho\pi$ do not couple to the strange vector current, the scenario (iv)
prediction for  $\langle r^2\rangle^s_D$ has a smaller magnitude and 
opposite sign to the scenario (i) and (ii) predictions. The assumption that 
$\bra{0}\bar{s}\gamma_\mu s\ket{\rho\pi}=0$, however, is inconsistent with the
phenomenology of $\phi$ decay, which displays a 13\% branch to
$\rho\pi$. To the extent that the $\rho\pi$ vector current form factors 
are $\phi$-meson dominated \cite{Goi96}, the $\rho\pi$ contribution to the 
nucleon isoscalar EM and strangeness moments should obey the relation in 
Eq. (\ref{resratphi}). Consequently, a more realistic scenario (iv) low-mass  
value for $\langle r^2\rangle^s_D$  should be closer to those from the 
other scenarios.

The remaining -- and dominant -- uncertainty is associated with the 
higher-mass content of the spectral functions. In the VMD approach, the 
effect of higher mass states is parameterized by a single $S'$ pole. Whether 
this pole represents a single resonance
with an in-principle well-defined flavor wavefunction or a sum over a tower 
of higher mass states ({\em viz}, $KK\pi$, $KK\pi\pi$...$\Lambda
\bar{\Lambda}$....) is uncertain.
Consequently, in the earlier works \cite{Jaf89,HMD96,For96} its 
contribution to the strange spectral function was fixed by requiring specific 
asymptotic behavior ($t\to\infty$) for the form factors. The choice of 
this condition is somewhat ambiguous, however, and the leading strange 
moments vary strongly for different 
reasonable choices for this condition \cite{For96,Mubos}.

We suggest an alternative method to quantify the uncertainty associated 
with the unknown higher mass spectral content. First, we argue that a 
purely hadronic description of the form factors is applicable only for 
relatively small momentum transfers ({\em e.g.}, 
$|t|\leq\mbox{ a few}$ (GeV$/c$)$^2$). From the standpoint of 
quark-hadron duality, we would expect a hadronic approach to produce 
the asymptotic $t$-dependence obtained from quark counting rules only when 
a sum over the entire hadronic spectrum is carried out. At present, 
performing this sum is not feasible. Consequently, a
hadronic framework should adequately describe the form factor only over 
a finite range of momentum transfer, as is used in the fits of the 
isocalar EM form factors.

Second, we assume that for each higher mass intermediate state, some 
relation exists between its contribution to the isoscalar EM and 
strangeneness spectral functions. Roughly speaking, the \lq\lq maximal" 
relation is given by Eq. (\ref{resratphi}):
whatever a state does in the isoscalar EM channel, it does about three 
times more strongly in the strangeness channel. States which do not 
contain resonating $s\bar{s}$ pairs would give relatively weaker 
contributions to the strangeness spectral function.
Using a single $S'$ pole to characterize the higher-mass spectral content, 
the largest
higher-mass effect would be given by assuming the ratio of its residues 
$(a_{S'}^{i})^s/s_{S'}^{i}$ is given by Eq. (\ref{resratphi}). 
We obtain a \lq\lq reasonable range" for the strangeness moments by 
adding and subtracting this maximal
$S'$ contribution to the low-mass values in Table 3.\footnote{Note that the 
case when the $S'$ is identified with the $\omega(1600)$ and couples 
according to Eq. (\ref{resratom}) is contained in this range.} 
Leaving the sign of the $S'$ uncertain allows for the possibility of 
cancellations between higher-mass contributions to the isoscalar 
EM spectral function which do not persist in the strangeness channel. 

In Table \ref{tab3} we show our results for scenarios (i), (ii), and (iv) 
(under the extreme assumptions for the $\bra{0}\bar{s}\gamma_\mu s
\ket{\rho\pi}$ discussed above).
\begin{table}
\begin{center}~
\begin{tabular}{|c|c||c|c|}\hline
Moment & Scenario & {\it low-mass value} & {\it reasonable range} 
\\\hline\hline
$\kappa^s $ & (i) & $-0.43$ & $-0.39\,\to\,-0.48$ \\
 & (ii) & $-0.28$ & $-0.15\,\to\,-0.41$ \\
 & (iv) & $-0.39$ & $-0.26\,\to\, -0.51$ \\\hline
$\langle r^2\rangle^s_D \ [\mbox{fm}^2]$ & (i) & $0.42$ & 
$0.42$ \\
 & (ii) & $0.42$ & $0.41\,\to\,0.43$ \\
 & (iv) & $-0.15$ & $-0.13\,\to\, -0.17$ 
\\\hline
\end{tabular}
\end{center}
\caption{\label{tab3} {\it Low-mass value} and {\it reasonable range}
for the leading strange moments $\kappa^s $ and $\langle r^2\rangle^s_D$
as defined in the text.}
\end{table}
We observe that both magnitude and sign for $\kappa^s$ are relatively robust. 
Differing assumptions for the higher mass contribution lead to at most 
a factor of 2-3 variation in the magnitude of $\kappa^s$ but no variation 
in the phase. It is difficult to generate a positive sign for $\kappa^s$ 
under any scenario and maintain consistency with the phenomenological 
constraints discussed earlier. The predictions for
the strange radius are less certain. Although the reasonable ranges 
for $\langle r^2 \rangle^s_D$ are generally smaller than those for 
$\kappa^s$, the low-mass value depends 
strongly on one's assumptions about the $3\pi\leftrightarrow\rho\pi$ 
coupling to $\bar{s}\gamma_\mu s$. We emphasize, however, that the result 
for scenario (iv) appearing in Table \ref{tab3} is likely to underestimate 
the magnitude of the radius, given the phenomenology of $\phi$ decays. 

Should the experimental value for either of the strangeness moments lie 
outside our reasonable range by more than one standard deviation, we would 
conclude that our treatment of the higher mass contributions is too 
na\"\i ve. In this respect, the first results for $\kappa^s$ from the 
SAMPLE collaboration \cite{Mue97} are suggestive.
It is conceivable, for example, that states such as the
$KK\pi$ could give resonance enhanced contributions to the strangeness 
form factors but only small effects in the isoscalar form factors 
(see {\em e.g.}, the initial studies of higher mass contributions 
in Ref. \cite{Bar98}). Such a scenario might arise from the
presence of two opposite sign pole contributions to the $KK\pi$ EM 
form factor ({\em e.g.}, $\omega(1600)$ and $\phi(1680)$) which do not 
both appear in the $KK\pi$ strangeness form
factor ({\em e.g.}, $\phi(1680)$ only).  At present, such scenarios 
remain speculative and await a more detailed analysis of the higher 
mass  strangeness spectral content. 

\section{Conclusions}
\label{sec:conc}
The spectral content of the isoscalar EM form factors appears to be 
considerably more complex than that of the isovector form factor. Only 
recently, for example, has the connection between continuum and resonant 
contributions to $\FOis$ and $\FTis$ been elucidated \cite{RMH98,HRM98}. 
In order to address some of the open questions regarding the isoscalar 
spectral content, we have drawn upon our previous study of the
$K\bar{K}$ continuum \cite{RMH98,HRM98} in the present reanalysis of the 
isoscalar form factors. The relative phase between the $K\bar{K}$ 
contribution and the 
vector meson poles has been determined from the electromagnetic fits.
We have then refitted the residues of the $\omega$, $\phi$, 
and higher mass ($S'$) poles in various scenarios in order to determine 
how much of the $\phi$ strength demanded by the data can be accounted for 
by the $K\bar{K}$-continuum. Our main findings are as follows:
\begin{itemize}
\item[(a)] the relative phase between $K\bar{K}$-continuum and
$\omega$ pole contribution is $0\,(\pi)$ for $\FO\,(\FT)$,
\item[(b)] the $\omega$ contributions to the form factors are stable
with respect to  differing treatments of other contributions, 
\item[(c)] for $\FTis$ all of the $\phi$ strength is accounted for 
by the $K\bar{K}$-continuum while for $\FOis$ additional $\phi$
strength (e.g. via $3\pi\to\rho\pi\to\phi$) is needed,
\item[(d)] the $K\bar{K}$ contribution to the form factors is $\phi$-resonance
enhanced yet is not well represented by a vector meson dominance approximation.
\end{itemize}

Furthermore, we have determined the $\phi NN$ coupling constants from the 
form factor fits and directly from the $K\bar{K}\to N\bar{N}$ partial waves.
The extracted coupling constants (cf. Tables \ref{tab1}, \ref{tab2})
imply large OZI-violation.
As observed previously for the $\omega NN$ couplings \cite{Dre98},
the results of the two methods do not agree. 
For a variety of reasons discussed above,
we deem it more sensible to extract the couplings directly from the 
partial waves.

We have also developed the implications for the leading
moments of the nucleon's strange vector form factors.
We quantify the main uncertainty in this
approach by splitting the leading moments into low-mass and  
higher mass contributions.
We quote a {\it low-mass value} that consists of
the reasonably well known $\omega$, $\phi$, and $K\bar{K}$ contributions.
These contributions are strongly constrained by isoscalar EM form factor data,
$KN$ partial waves, $e^+e^-\to K\bar{K}$ cross sections, and vector meson octet
phenomenology.  We also give a {\it reasonable range} by considering the
scenarios in which the remaining higher-mass intermediate state contributions,
parameterized by a single vector meson pole
$S'$, couples to strangeness maximally like the $\phi$ (or with an 
opposite sign). We find that both the magnitude and negative sign 
for $\kappa^s$ are
rather robust for various scenarios, whereas the predictions for $\langle
r^2\rangle^s_D$ contain more variation. 
Whether there exist additional higher-mass contributions which
would modify our reasonable ranges for the strangeness moments yet 
which do not affect the isoscalar EM form factors remains to be seen.

\section*{Acknowledgement}
We wish to thank T. Cohen, R.L. Jaffe, and U.-G. Mei{\ss}ner
for useful discussions. HWH acknowledges the hospitality of the Institute
for Nuclear Theory in Seattle where part of this work was carried out.
MJR-M has been supported in part under U.S. Department of Energy contract
\# DE-FG06-90ER40561 and under a National Science Foundation Young
Investigator Award. HWH has been supported by the Natural Sciences
and Engineering Research Council of Canada.

\end{document}